# A bimetal and electret-based converter for thermal energy harvesting

Sebastien Boisseau, Ghislain Despesse, Stephane Monfray, Onoriu Puscasu and Thomas Skotnicki

*Abstract*— **This paper presents a new device able to turn thermal gradients into electricity by using a bimetal-based heat engine coupled to an electrostatic converter. A two-steps conversion is performed: (i) a curved bimetallic strip turns the thermal gradient into a mechanical movement (thermal-to-mechanical conversion) that is (ii) then converted into electricity thanks to an electret-based electrostatic converter (mechanical-to-electrical conversion). An output power up to 5.5µW on a hot source at 50°C has already been reached, validating this new concept.**

*Index Terms*—**Electrets, electrostatic devices, thermoelectric energy conversion, transducers.**

## I. INTRODUCTION

SMALL-SCALE energy harvesting (EH) is a field of growing interest with a market estimated at more than $4.4 billion within ten years [1]. Many principles of EH have been investigated [2], and among them, thermal EH has proven attractive especially when solar EH is not possible. In fact, many small-scale thermal energy harvesters have already been developed using bimetallic junctions that generate a thermoelectric voltage (Seebeck effect) when they are submitted to a temperature gradient; besides, some of them are commercialized for a long time. These devices have good conversion efficiencies (10-15% of Carnot efficiency), but are quite expensive due to the cost of the materials (rare-earth materials such as bismuth telluride).

We present hereafter an alternative concept that uses a curved bimetallic strip inserted into an electret-based electrostatic converter and enabling to turn thermal gradients into electricity as well. This conversion has the strong advantage, compared to standard thermoelectric energy harvesters, to be a low cost solution as it employs standard and fully available materials.

## II. BIMETAL AND ELECTRET-BASED CONVERTERS FOR THERMAL ENERGY HARVESTING

Bimetals are made of two strips of different metals with different coefficients of thermal expansion (CTE) (e.g. iron and copper) that are joined together. This difference in CTE

S. Boisseau and G. Despesse are with the CEA, LETI, Minatec Campus, 17 avenue des martyrs, Grenoble, France. (e-mail: sebastien.boisseau@cea.fr)
S. Monfray, T. Skotnicki and O. Puscasu are with ST Microelectronics, 850 rue J. Monnet, Crolles, France.

enables flat bimetallic strips to bend when heated up or cooled down (figure 1a). Curved or stamped bimetallic strips (figure 1b) are even smarter devices, presenting strong nonlinear behaviors, and able to snap and snap-back between two positions (sudden buckling) according to the temperature with a hysteretic behavior (figure 1c). This phenomenon has been thoroughly studied by Timoshenko [4] and others [5-8]. Today, curved bimetals are used in many electrical and mechanical devices [4,9-10] such as actuators, clocks, thermometers, thermostats, circuit breakers, time-delay relays….

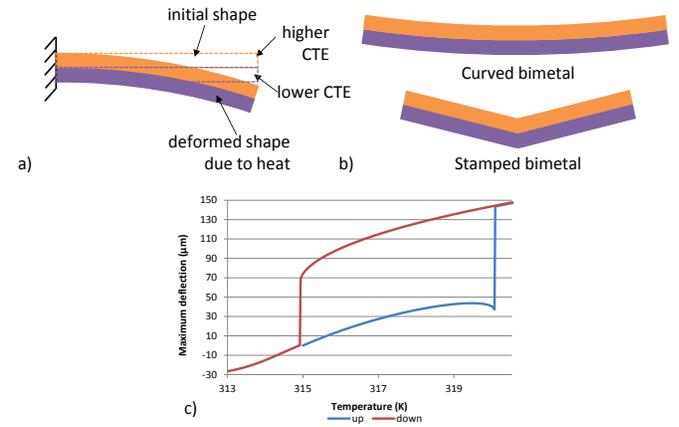

Fig. 1. (a) flat bimetallic strip, (b) curved bimetallic strip and (c) hysteretic cycle as a function of the temperature for a curved bimetal

Here, a curved bimetal is used as a heat engine able to turn a thermal gradient into a mechanical movement. The bimetal is clamped in a cavity with a hot source on the bottom and a cold source (e.g. ambient air, heat sink, cold element…) on the top as presented in figure 2a. The metal with the higher CTE (noted CTE+) is above the metal with the lower CTE (noted CTE-) in this configuration.

At the equilibrium temperature (e.g. T=25°C), the device is in position 1 (figure 2a): the bimetal is in contact with the lower plate. The device is then placed on a hot source; the bimetal is heated up, accumulates mechanical elastic energy, suddenly snaps to position 2 (figure 2b) and enters in contact with the upper plate. There, it is cooled down and snaps back to position 1; and a new cycle restarts. As a consequence, the bimetal oscillates between two states and the thermal gradient is turned into a mechanical movement. A piezoelectric-based device using this concept has already been proposed by [11]. In this paper, an electret-based converter has been chosen to turn the mechanical movement into electricity.

Electrets are dielectrics capable of keeping electric charges for years. These materials are the equivalent to magnets in electrostatics, and have already proven their benefits in many fields such as transducers (microphones) [12], microfluidics [12], actuators (micromotors) [12] and energy harvesting [13-15] as a high voltage polarization source.

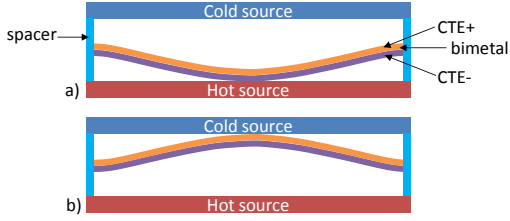

Fig. 2. (a) Bimetal-based heat engine in position 1 and (b) in position 2 after snapping.

Electret-based converters are electrostatic converters, and are therefore based on a capacitive structure made of two plates (electrode and counter-electrode) (figure 3a). The electret induces charges on the electrode and on the counter-electrode to respect Gauss's law. $Q_t$, the charge on the electret is equal to the sum of $Q_1$ and $Q_2$, where $Q_1$ is the total amount of charges on the electrode and $Q_2$ the total amount of charges on the counter-electrode ($Q_t = Q_1 + Q_2$). A relative movement of the counter-electrode compared to the electret and the electrode induces a change in the capacitor shape (e.g. the counter-electrode moves away from the electret, changing the air gap and then the electret's influence on the counter-electrode) and leads to a reorganization of charges between the electrode and the counter-electrode through load $R$. This results in a current circulation through $R$: the mechanical movement is turned into electricity. The equivalent model of this converter is a voltage source in series with a variable capacitor as presented in figure 3b and is ruled by the differential equation $dQ_2/dt = V_s/R - Q_2/(C(t)R)$, where $V_s$ is the electret's surface voltage and $C(t)$ the converter's capacitance (between the electrode and the counter-electrode).

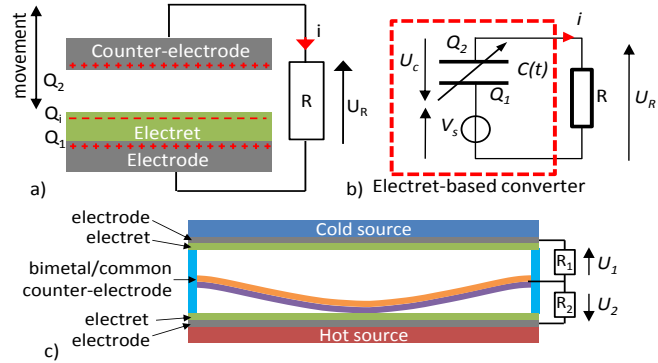

Fig. 3. (a) Electret-based converter, (b) equivalent electrical model and (c) electret-based converter coupled to the bimetal-based heat engine

In fact, the heat engine presented in figure 2 can be easily coupled to the electret-based converter by adding 2 electret layers on the upper and the lower plates and by using the bimetal as the moving electrode (figure 3c). Two variable capacitors are then formed and can be exploited to turn the mechanical energy from the bimetal into electricity. Then, electric power can be harvested both on $R_1$ (channel 1) and $R_2$ (channel 2).

## III. EXPERIMENTAL RESULTS AND DISCUSSION

To validate this concept, a prototype (figure 4a) has been made from two plates of oxidized silicon (upper and lower plates). The 2μm-thick $SiO_2$ layer, covered by a 100nm-thick parylene layer and charged by a standard triode corona discharge (point-grid configuration), is used as the electret. A 115μm-thick curved bimetal with a weight of 0.26g and a surface of 1cm×3cm, designed to snap at 47°C and to snap-back at 42.5°C, is inserted into the 515μm-thick cavity between the two plates. The bimetal is in B72M (Mn-Cu18%-Ni10%) that has a high CTE ($\alpha=26.4\times10^{-6}$) and in INVAR (Fe-Ni36%) that presents a very low CTE ($\alpha=2\times10^{-6}$), and is coated with a 100nm-thick parylene layer to protect the electret's surface charges during contacts. The other dimensions are shown in figure 4a.

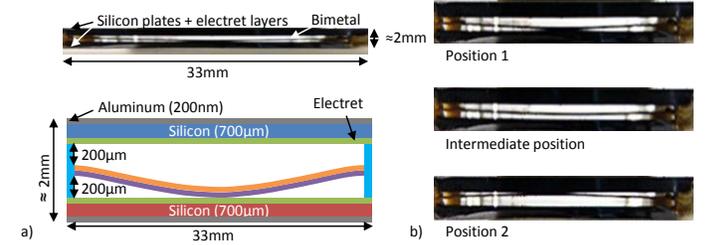

Fig. 4. (a) Prototype and dimensions and (b) zoom on the bimetal for position 1, position 2 and intermediate position.

The device is placed on a hot source at 50°C; the upper plate is cooled by air. As expected, the bimetal snaps through and back and therefore, oscillates between the two plates. Some positions of the bimetal are shown in figure 4b (for better clarity, the photo is shrunk in the horizontal direction). The output voltages on a 1GΩ-load on both channels and for an electret's surface voltage of $V_s$=240V are presented in figure 5. The snapping frequency is about 1Hz and like all electret-based converters, output voltages with high impedances reach several hundreds of volts. The maximum voltages are reached when the bimetal contacts the electret. Instantaneous output power reaches 341μW on channel 1, but the average output power of the device is only 5.5μW due to the low snapping frequency.

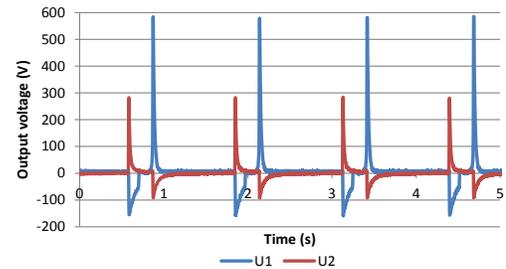

Fig. 5. Output voltages on a 1GΩ load (channel 1 and channel 2)

Moreover, the electret-based converter has proven to be particularly suitable for this device, enabling to get high electromechanical couplings. Indeed, the electret-based converter has a strong impact on the bimetal's snapping frequency: the higher the surface voltage of the electret, the higher the output power and the lower the snapping frequency becomes, as presented in figure 6. And finally, the electret-based converter is strong enough to divide the bimetal's snapping frequency by two when it harvests 5.5μW, validating the high electromechanical couplings.

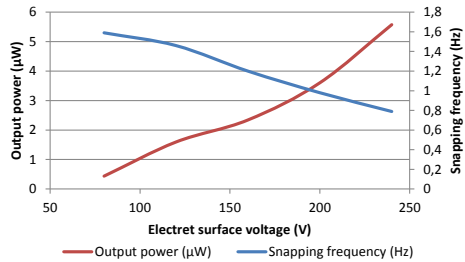

Fig. 6. Impact of the electret's surface voltage on the snapping frequency – output powers and snapping frequency as a function of the electret's surface voltage

## IV. CONCLUSION

We have developed a new concept of thermal energy harvester based on a bimetal heat engine and an electret-based converter. An average output power up to 5.5μW has been reached on a hot source at 50°C and with an ambient temperature of 25°C. These devices pave the way to new low-cost thermal energy harvesters made with simple and fully available materials (copper, iron, $SiO_2$…).

## ACKNOWLEDGMENTS

The authors would like to thank G. Pitone, Delta Concept, for the design of the bimetals.